\documentclass[reprint,amsmath,amssymb,aps,prl,superscriptaddress]{revtex4-2}

\usepackage{graphicx}
\usepackage{bm}
\usepackage{dsfont}

\newcommand{\E}{\ensuremath{\mathbb{E}}}
\newcommand{\D}{\ensuremath{\mathcal{D}}}
\newcommand{\rd}{\,\mathrm{d}}
\newcommand{\Var}{\mathrm{Var}}
\newcommand{\vol}{\ensuremath{|\D|}}

\begin{document}

\title{Statistical mechanics of spontaneous behavior}

\author{Daniel Yasumasa Takahashi}
\email{takahashiyd@gmail.com}
\affiliation{Brain Institute, Federal University of Rio Grande do Norte, Natal, Brazil}
\author{Ahmed El Hady}
\email{ahmed.el-hady@uni-konstanz.de}
\affiliation{Centre for the Advanced Study of Collective Behaviour, University of Konstanz, Konstanz, Germany}
\affiliation{Department of Collective Behaviour, Max Planck Institute of Animal Behavior, Konstanz, Germany}

\date{\today}

\begin{abstract}
Why do a few behaviors fill most of an animal's time? The usual answer is that these
behaviors are started more often, because learning and reinforcement favor them. Here we
show that how long a behavior lasts can, by itself, explain much of what is observed. We
consider a system that runs until it meets a constraint, such as a wall, and then resets.
The fraction of time it spends in each behavior then follows the Boltzmann law, and the
partition function is the mean time between resets. The four laws of thermodynamics follow
as theorems from four postulates, and critical points that act as bottlenecks become the most occupied states.
In 47 multi-day recordings of fruit flies, the dominant behavior accounts for only $13\%$ of
bout starts yet fills $49\%$ of the time, and its typical duration identifies the individual
fly. These results lay the groundwork for a statistical mechanics of behavior.
\end{abstract}

\maketitle

Place a fly in a closed box and record what it does. The record has a recognizable shape.
Long stretches of circling alternate with short crossings and rare collisions with a wall,
and of everything the fly could do, a few behaviors take up most of its time. The same is
true of any agent whose activity is interrupted by encounters with a constraint, whether
that constraint is a wall, the limit of its own body, or an exhausted energy reserve.

Two quantities summarize such a record. The \emph{frequency} of a behavior is how often it
is started, and its \emph{duration} is how long it lasts once started. The time a behavior
fills is the product of the two, so a behavior can dominate because it is started often or
because it lasts long. The main theories of behavior, such as reinforcement learning and
choice theory, explain dominance through frequency. They ask which action is selected and
treat its duration as a detail.

Here we ask what a record looks like when nothing favors one behavior over another, so that
only duration varies. The answer is a \emph{null theory}, which plays the role that the
ideal gas plays in thermodynamics. We show that this null theory is a complete statistical
mechanics. The time share of each behavior follows the Boltzmann law, the partition function
is the mean time between two encounters with the constraint, and the four laws of
thermodynamics follow as theorems from four postulates. Dominance, the matching law,
individuality, and criticality all emerge without assigning value to any behavior. Formally
the construction belongs to stochastic resetting~\cite{EvansMajumdar2011,EvansMajumdarSchehr2020},
with one difference that matters. The reset is triggered when the system reaches the
constraint, and this is what makes the partition function a time. We state the postulates,
derive the free energy and the four laws, and test the theory in three simulated systems and
in 47 multi-day fly recordings.

\paragraph*{The postulates.}
\emph{P0 (behavior).} An observer chooses a finite set of observables, such as position and
heading, and a time resolution. The behavior of a system over a record is then its path
$t\mapsto X_t$ through the space $\mathcal X$ that these observables
span~\cite{Berman2014,BrownDeBivort2018}. Because the observables are a choice, one animal
supports many behaviors, and the laws below hold for each. The dynamics of $X_t$ depends on
a hidden parameter $a\in\mathcal A$, which we call an \emph{affordance}, after Gibson's term
for what a situation offers an agent to do~\cite{Gibson1979}. Each value of $a$ fixes one
regime of the dynamics, for example circling or crossing. We also call it a \emph{mode}.
Finally, a region $\D\subset\mathcal X$ is bounded by the \emph{constraint} $C=\partial\D$.
The constraint need not be a wall. A joint at the end of its range or a depleted energy
store ends a bout in the same way.

\begin{figure*}[t]
\centering
\includegraphics[width=0.9\textwidth]{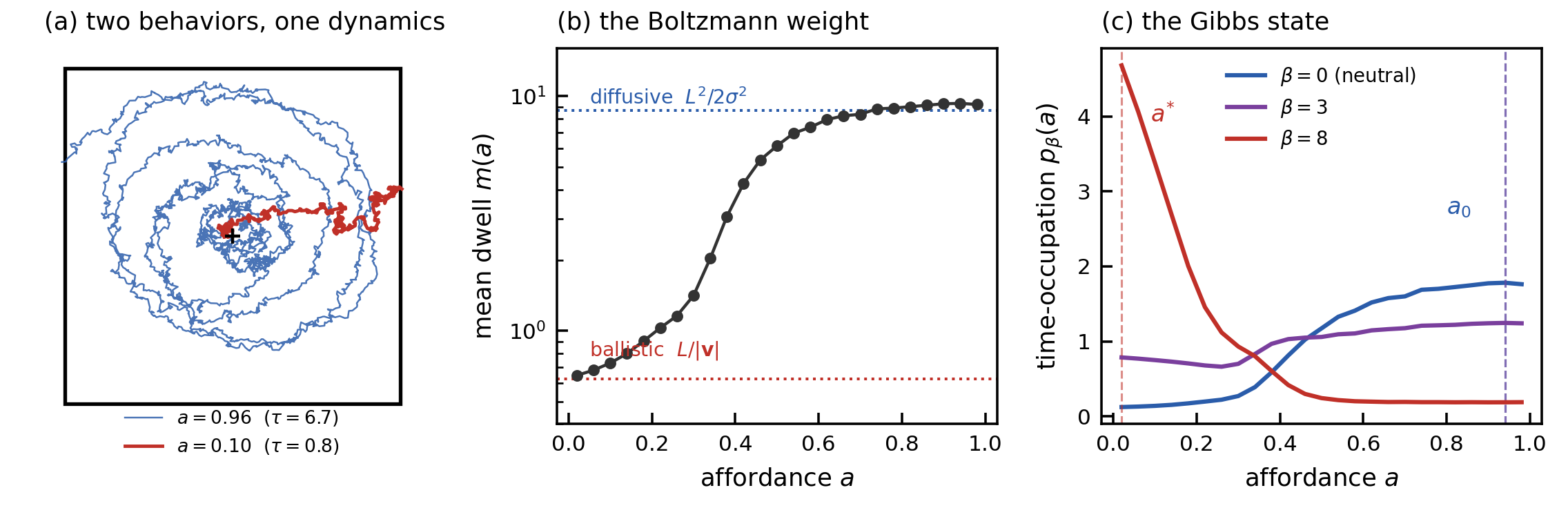}
\caption{The box, Eq.~\eqref{eq:box}. (a)~Two behaviors from one dynamics. Circling
($a=0.96$, blue) is trapped and drift ($a=0.10$, red) crosses.
(b)~The dwell $m(a)$ rises from the ballistic crossing time $0.65\approx L/|\mathbf v|$ to the
diffusive escape time $9.2\approx L^2/2\sigma^2$.
(c)~Occupation $p_\beta(a)$, where $\beta$ is the strength of the agent's response to reward
(see \emph{Reward}). With no reward ($\beta=0$) and a uniform frequency the record is
dominated by circling through duration alone. As reward for reaching the wall grows
($\beta=3,8$) drift takes over. Parameters $L=1$, $\omega=6$, $|\mathbf v|=1.6$,
$\sigma=0.24$, over $\sim8\times10^5$ excursions.}
\label{fig:box}
\end{figure*}

\emph{P1 (reset).} With the affordance $a$ held fixed, the observable follows a memoryless
(Markov) dynamics until its path reaches the constraint $C$. At that contact two things
happen. The observable restarts from a fixed starting distribution, the \emph{injection
law}, and a new affordance is drawn from the \emph{resampling law} $\lambda$. We call this an
\emph{affordance-reset} process. Since $\lambda$ sets how often each affordance is started,
it is the frequency of the theory. The stretch between two contacts is an
\emph{excursion}. Its length $\tau^a=\inf\{t:X_t\in C\}$ is the duration, and its mean
$m(a)=\E[\tau^a]$ is the \emph{dwell} of $a$. Our reference system throughout is a diffusion
in a box,
\begin{equation}
\rd\mathbf X_t=\big[a\,\omega(-Y_t,X_t)+(1-a)\,\mathbf v\big]\rd t+\sigma\rd\mathbf W_t,
\label{eq:box}
\end{equation}
where $a=1$ is pure rotation, which the geometry traps, and $a=0$ is pure drift, which
crosses the box ballistically [Fig.~\ref{fig:box}(a)]. Between the two the dwell rises by
more than an order of magnitude [Fig.~\ref{fig:box}(b)].

\emph{P2 (the ensemble is time).} Ordinary statistical mechanics averages over many copies
of a system. Here there is one record and no copies, so the only primitive quantity is the
\emph{occupation}, the fraction of a window of length $t_w$ that each mode fills,
$\hat p_{t_w}(\rd a)=t_w^{-1}\int_0^{t_w}\mathds 1(a_{t'}\in\rd a)\,\rd t'$. Equilibrium means
that this fraction has a limit as $t_w\to\infty$. That limit plays the role of the
thermodynamic limit, so in this theory \emph{system size is recording duration}. Under P1
successive excursions are independent and form a renewal process, and the renewal reward
theorem~\cite{Cinlar1975} gives the long-run occupation,
\begin{equation}
p_0(\rd a)=\frac{m(a)\,\lambda(\rd a)}{Z_0},\qquad Z_0=\int m\,\rd\lambda=\langle\tau\rangle_\lambda.
\label{eq:gibbs}
\end{equation}
In words, the time share of a mode is its frequency times its dwell, divided by the mean
duration of an excursion. This is a Gibbs measure written in the two variables of the
record. The frequency $\lambda$ plays the role of the density of states, the dwell $m$ that
of the Boltzmann weight, and the partition function $Z_0$ is the \emph{mean cycle time}.
Equation~\eqref{eq:gibbs} treats all modes alike. A mode that holds the system longer is
simply observed for longer, an effect statisticians call length-biased sampling. This is why
circling dominates the box even when every mode is started equally often
[Fig.~\ref{fig:box}(c), blue]. The agreement of time and ensemble averages, which ordinary
statistical mechanics must assume, is here a theorem.

P2 also limits what a finite record can reveal. A window of length $t_w$ contains about
$N=t_w/Z_0$ excursions, and the occupation measured in it fluctuates around $p_0$ with
variance $\chi/N$, where $\chi$ is given in the SM. The number of excursions thus plays the
part of the number of particles in a finite gas.

\emph{P3 (the dictionary).} It remains to name temperature and energy. For a diffusion in a
potential landscape with noise strength $\sigma$, the exit theory of Freidlin and
Wentzell~\cite{FreidlinWentzell2012} shows that the dwell grows exponentially in
$1/\sigma^2$, $\sigma^2\log\E^\sigma[\tau^a]\to\bar J^a$, where $\bar J^a$ is twice the
barrier to the easiest exit. Because $\bar J^a$ is a property of the landscape and does not
depend on the noise, the natural way to write the dwell as a Boltzmann factor is
\begin{equation}
T:=\sigma^2,\qquad E(a):=-\bar J^a,\qquad m(a)=e^{-E(a)/T}.
\label{eq:dict}
\end{equation}
The energy of a behavior is therefore minus its exit barrier, so a deep trap is a
low-energy state. The environment sets the energies, while the temperature (the agent's
noise) and the frequency belong to the agent.

\paragraph*{The potential.}
As in thermodynamics, the theory is organized around a free energy and its derivatives. The
free energy of one excursion is $F=-T\log(Z_0/\tau_0)$, with $\tau_0$ a reference time, and a
record of $N$ excursions has total free energy $\Phi=NF$, so $N$ plays the role of system
size, with conjugate $F=\partial\Phi/\partial N$. The free energy $F$ depends on four
controls. These are the temperature $T$, the volume $\vol$ of the domain $\D$, the strength
$\beta$ of a reward field introduced below, and the propensities $\mu_i$. The propensities
are the shares of the frequency given to groups of behaviors, with $\sum_i\mu_i=1$. Each
control has a \emph{conjugate}, the partial derivative of $F$ with respect to that control
with the others held fixed,
\begin{equation}
\rd F=-S\,\rd T-\Pi\,\rd\vol-\Upsilon\,\rd\beta+\sum_i\mathrm v_i\,\rd\mu_i .
\label{eq:diff}
\end{equation}
All conjugates are per excursion, and those of the whole record follow by multiplying by
$N$. Each has a behavioral reading. The entropy $S=-\partial F/\partial T$ measures how far
the occupation departs from the frequency. It equals minus the relative entropy
$-\!\int p_0\log(p_0/\lambda)$, so it vanishes when duration plays no role. The pressure
$\Pi=-\partial F/\partial\vol$ is the response of the free energy to enlarging the domain.
The mean reward $\Upsilon=-\partial F/\partial\beta$ is $T$ times the average reward of the
record. The behavioral chemical potential $\mathrm v_i=\partial F/\partial\mu_i$ is the
free-energy cost of shifting frequency into group $i$. Proofs are in the SM.

\paragraph*{The four laws.}
Each law is a theorem of P0 to P3 (Table~\ref{tab:laws} in the SM). For each one we state
what frequency can do and what duration does.

\begin{figure*}[t]
\centering
\includegraphics[width=0.9\textwidth]{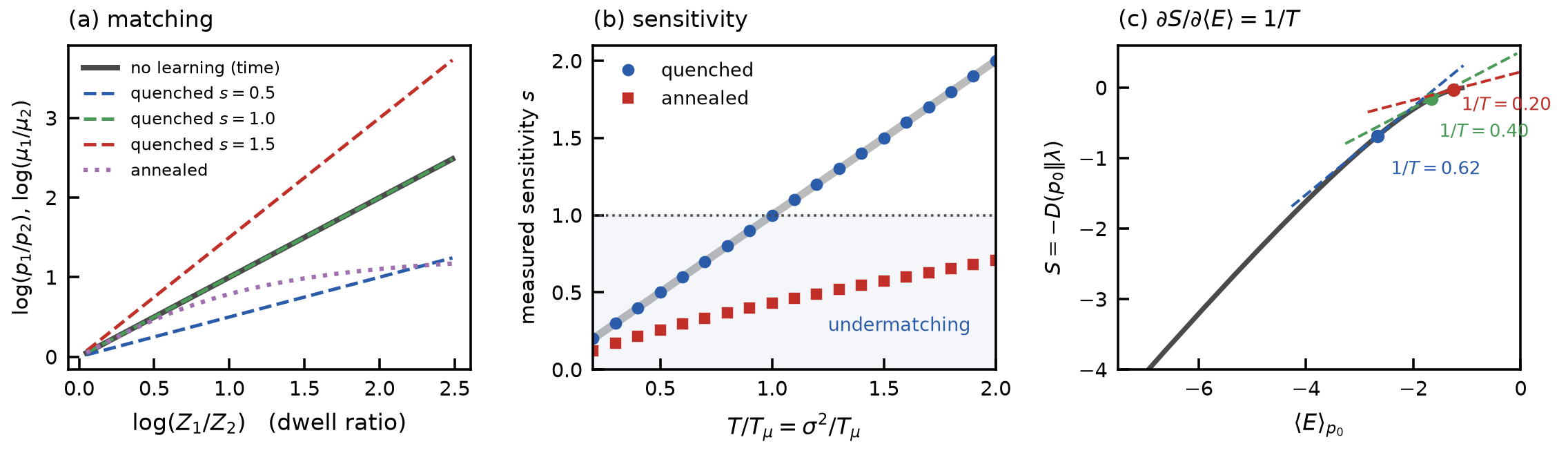}
\caption{Zeroth and first laws. (a)~Log ratios against the dwell ratio of two options.
Time shares $p_1/p_2$ with uniform propensity give strict matching [Eq.~\eqref{eq:alloc},
heavy line]. Propensities $\mu_1/\mu_2$ after exchange follow the generalized matching law
$\mu_i\propto Z_i^{\,s}$ when the groups are kept separate (quenched, dashed) and bend when
they are pooled (annealed, dotted) [SM].
(b)~The sensitivity $s$ measured from the slope in (a) against $T/T_\mu$
[Eq.~\eqref{eq:sens}]. Undermatching ($s<1$, shaded) is the regime $T<T_\mu$. (c)~The
entropy against the mean energy, for the trap ensemble with $T_\lambda=1$, is concave with
slope $1/T$, as the first law requires.}
\label{fig:alloc}
\end{figure*}

\emph{Zeroth law} (a common scale exists). We first need the time allocated among options.
Split the repertoire into disjoint groups $A_i$, for example the options of a choice task.
Let $\mu_i$ be the propensity of $A_i$, the share of fresh affordances drawn from it, and let
$Z_i=\int_{A_i}m\,\rd\lambda_i$ be the mean dwell of a visit to it, with $\lambda_i$ the
frequency within $A_i$. Since the agent enters
$A_i$ at a rate proportional to $\mu_i$ and each visit lasts $Z_i$ on average, the renewal
argument behind Eq.~\eqref{eq:gibbs} gives
\begin{equation}
p_i=\frac{\mu_i Z_i}{\sum_j\mu_j Z_j}.
\label{eq:alloc}
\end{equation}
With \emph{uniform} propensities, that is with no preference, reward, or learning, this
reduces to $p_i\propto Z_i$. The time allocated to an option is then proportional to its
dwell, with slope one [Fig.~\ref{fig:alloc}(a), heavy line]. This is the matching law of
operant psychology~\cite{Herrnstein1961,Baum1974}, the central law of choice, here produced
by duration alone.

The zeroth law itself concerns what happens when propensity can flow between groups. The
chemical potential $\mathrm v_i$ is the cost of moving a little frequency into $A_i$, and at
equilibrium it is equal across groups. Because it is a single number, this equality is
transitive. If $A_1$ is in equilibrium with $A_2$ and $A_2$ with $A_3$, then $A_1$ is in
equilibrium with $A_3$, and $\mathrm v$ serves as a common scale, as temperature does in the
classical law. If allocating propensity carries a mixing entropy at a temperature $T_\mu$,
equalizing $\mathrm v$ redistributes propensity as $\mu_i\propto Z_i^{\,s}$, the
generalized matching law, with sensitivity
\begin{equation}
s=\frac{T}{T_\mu},
\label{eq:sens}
\end{equation}
where $T_\mu$ measures how strongly the agent resists concentrating its propensity on a
few groups [SM]. The theory therefore predicts undermatching, the
empirically common case (median $s=0.83$ for responses and $0.92$ for time in Baum's
survey~\cite{Baum1979,WeardenBurgess1982}), whenever the agent's noise $T$ is smaller than
$T_\mu$ [Fig.~\ref{fig:alloc}(a,b), shaded].

\emph{First law} (heat, work, and a state function). Let $U=\langle E\rangle_{p_0}$ be the
mean energy of the record, that is, the time average of minus the exit barrier of the
current mode. A record that spends its time in deep traps has low $U$. When a control such
as the noise $T$ or the box size $\vol$ is changed slowly, $U$ changes for two reasons.
Either the record redistributes its time among the modes while their energies stay fixed,
or the environment shifts the energies while the time shares stay fixed. The first change
is heat, and for a Gibbs occupation it equals exactly $T\rd S$ [SM]. The second is work,
and its volume part is $-\Pi\rd\vol$,
\begin{equation}
\rd U=\underbrace{\textstyle\int E\,\rd p_0}_{\delta Q\,=\,T\rd S}
\;+\;\underbrace{\textstyle\int p_0\,\rd E}_{\delta W\,=\,-\Pi\,\rd\vol}.
\label{eq:first}
\end{equation}
Because the environment sets the energies directly, the law here carries no conservation
statement. Its content is the decomposition itself. Every change of $U$ is heat
or work, and every conjugate in Eq.~\eqref{eq:diff} derives from the single function
$F$. This has testable consequences. The pressure measured from the response to volume
must agree with the one computed from the mean cycle time, and the mixed derivatives must
obey Maxwell relations such as $\partial\Pi/\partial\beta=\partial\Upsilon/\partial\vol$
[SM]. The entropy is concave in the energy with slope exactly $1/T$
[Fig.~\ref{fig:alloc}(c)]. The law requires a reward that does not depend on the box
geometry, because otherwise the force measured from the geometry is not a derivative of $F$.

\emph{Second law} (two irreversibilities). The second law appears in two forms. The first
concerns the approach to equilibrium, and the second concerns the equilibrium record
itself. Relaxation \emph{toward} the Gibbs occupation is monotonic. For any resampling rule
with stationary law $\rho$, the distribution $q_n$ of the mode after $n$ excursions obeys
$D(q_{n+1}\Vert\rho)\le D(q_n\Vert\rho)$, so the distance to equilibrium can only decrease, an
$H$-theorem~\cite{CoverThomas2006} [SM]. The equilibrium record also has a direction in
time, and it enters at the reset. At each contact many different states are sent back to
the same injection point, and this map cannot be run backward. The observed sequence is
therefore irreversible even when the affordance is resampled symmetrically. Its entropy
production is the occupation average of $\dot S_{\rm exc}(a)$, the rate at which excursions
in mode $a$ produce entropy through their reset,
\begin{equation}
\dot S_{\rm obs}=\big\langle\dot S_{\rm exc}\big\rangle_{p_0}\ \geq\ 0,
\label{eq:second}
\end{equation}
with equality only when there is no reset~\cite{Seifert2012,Crooks1999}. In data the
irreversibility $\mathcal I$, the relative entropy between the statistics of the observed
transitions and those of the same record played backward, gives a measurable lower bound on
it [SM].

\emph{Third law} (unattainability). The classical third law says that absolute zero cannot
be reached in finite time. Its counterpart here is that no record can be completely
dominated by one behavior in finite time. Let $\mathcal G$ be a set of modes that are
candidates to dominate, and suppose the other modes are still started, with frequency
$\lambda(\mathcal G^c)>0$. Every excursion lasts at least the fastest crossing of the domain,
$\tau_{\min}$, so each cycle spends on average at least $\tau_{\min}\lambda(\mathcal G^c)$
outside $\mathcal G$. Dividing by the mean cycle time gives
\begin{equation}
Z_0\ \ge\ \frac{\tau_{\min}\,\lambda(\mathcal G^c)}{p_0(\mathcal G^c)}.
\label{eq:third}
\end{equation}
Pushing the time share outside $\mathcal G$ toward zero therefore requires the mean cycle
time to grow without limit. The bound shows how differently the two variables act.
Frequency is bounded, $\lambda\le1$, so it can make one behavior fill the record only by
never starting the others. Duration is unbounded, so it can make one behavior fill the
record with the repertoire intact, but only by making excursions ever longer. A record of length $t_w$ contains $N=t_w/Z_0$ excursions
and knows its own occupation only to a scatter $\chi/N$ [P2]. As the occupation
concentrates, $Z_0$ grows, $N$ falls below one, and the window becomes a single unfinished
excursion. No finite record ever shows a fully dominated occupation. In classical
thermodynamics this unattainability is an independent principle, or requires additional
resource assumptions~\cite{MasanesOppenheim2017}. Here it is a theorem, because the
partition function is itself a time.

The limit itself is still informative. As the noise $T$ goes to zero the occupation
collapses onto the mode with the deepest exit barrier, $\mathcal G=\arg\max_a\bar J^a$, which
is the longest-dwelling mode, and the entropy tends to $\log\lambda(\mathcal G)$ [SM]. A
record concentrated on its most persistent behavior looks optimized, yet nothing has been
optimized. A Gibbs measure has simply concentrated on its energy minimum, as a cooled gas
does, and the recording time needed to observe this grows with $Z_0$ without bound.

\begin{figure}[b]
\centering
\includegraphics[width=\columnwidth]{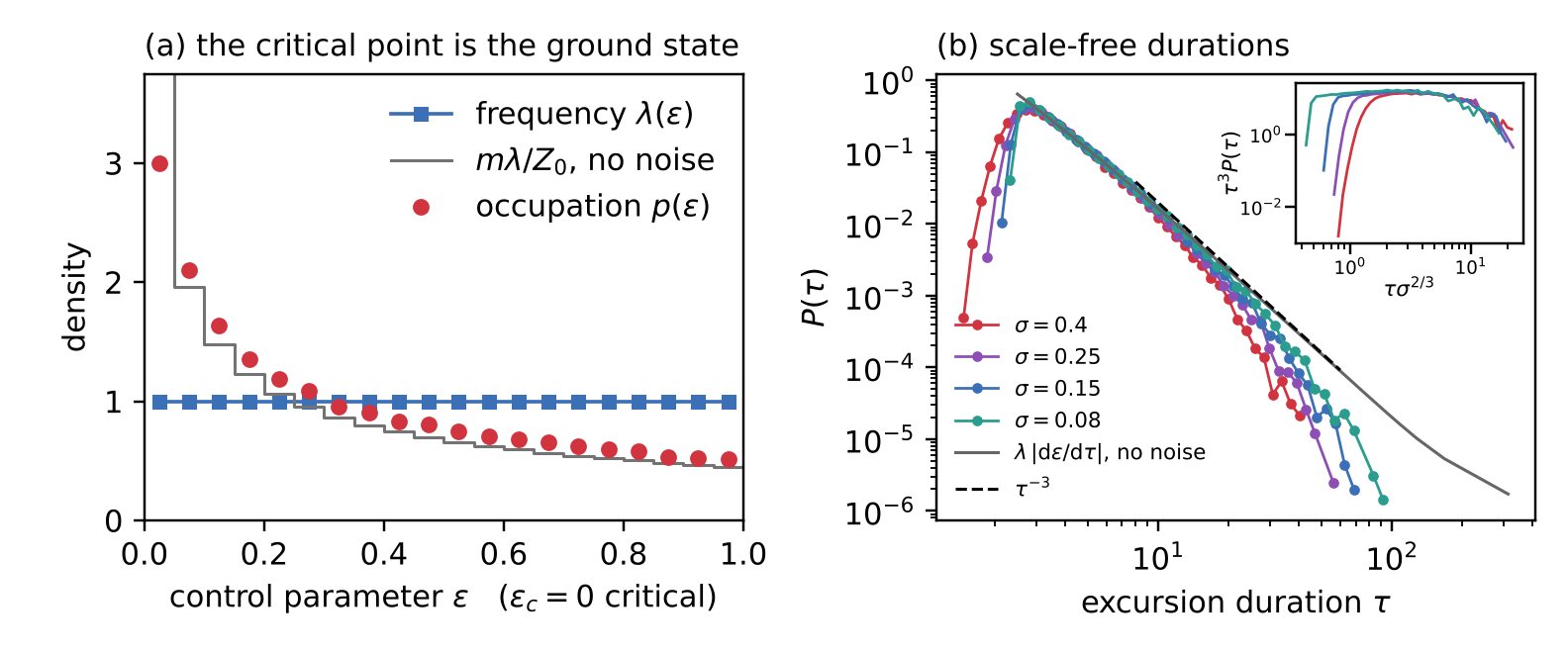}
\caption{The critical point is the ground state. Saddle-node bottleneck, critical at
$\epsilon_c=0$. (a)~From a uniform frequency $\lambda(\epsilon)$ (blue) the occupation
$p(\epsilon)\propto m\lambda$ (red) piles up at the critical point. (b)~Excursion durations
are scale-free, $P(\tau)\sim\tau^{-3}$ (dashed), cut off by noise. Inset, the collapse
$\tau^3P(\tau)$ against $\tau\sigma^{2/3}$. $X_0=3$, uniform $\epsilon$ on $(0,1]$,
$\sigma=0.15$ in (a).}
\label{fig:soc}
\end{figure}

\emph{Criticality.} The energy minimum has a further identity. When an excursion must pass through a bottleneck
near a continuous transition, the passage time diverges, the phenomenon called critical
slowing down. The critical point is then the longest-dwelling affordance, so it minimizes
$E=-T\log m$ and is the ground state. Duration alone parks the record at the critical point.
Figure~\ref{fig:soc} shows the simplest case, a saddle-node bottleneck
$\rd x=(\epsilon+x^2)\rd t+\sigma\rd W_t$, reinjected at $-X_0$ on reaching $X_0$, in which
the control parameter $\epsilon$ is the affordance and $\epsilon_c=0$ is critical. The
passage time grows as $\pi/\sqrt\epsilon$, so from a flat frequency the occupation piles up
at $\epsilon_c$ as $\epsilon^{-1/2}$ [Fig.~\ref{fig:soc}(a)]. Because each $\epsilon$ maps to
one duration, the durations follow a power law, $P(\tau)\sim\tau^{-3}$, up to a cutoff set by
the noise, $\tau^\ast\sim\sigma^{-2/3}$ [Fig.~\ref{fig:soc}(b)]. Observed criticality is
usually read as evidence of tuning or of a self-organizing
feedback~\cite{BakTangWiesenfeld1987,Munoz2018}. Here it is the null expectation for any
repertoire that straddles such a transition, and it must be excluded before criticality is
attributed to adaptation.

A power law of durations is the signature of a heavy-tailed repertoire, where the third law
takes a sharper form. Suppose the frequency itself has a Boltzmann form in the exit
barriers, $\lambda(\bar J)\propto e^{-\bar J/T_\lambda}$, with $T_\lambda$ setting how fast
the frequency decays with barrier depth. This is the trap ensemble of
Bouchaud~\cite{Bouchaud1992,MonthusBouchaud1996}. The partition function is then
$Z_0=T/(T-T_\lambda)$ [SM]. It is finite when the noise $T$ exceeds $T_\lambda$ and diverges
at $T=T_\lambda$, a point we call the \emph{ergodic edge}, because there time averages stop
converging. Beyond the edge, $T<T_\lambda$, the number of excursions in a window grows more
slowly than the window, $N(t_w)\sim t_w^{\alpha}$ with $\alpha=T/T_\lambda<1$. As in the trap
model, the occupation of a window is then expected to remain random at all times, which is
called weak ergodicity breaking, and the statistics of a record to change with its length,
which is called aging~\cite{Bouchaud1992,MonthusBouchaud1996}. The exponent $\alpha$ can be
read off the tail of the measured durations, so the duration exponent of a record tells how
far it is from the edge.

\paragraph*{Reward.}
So far every mode has been treated alike. Now let each mode $a$ return a reward $\Psi(a)$
when it is run, and let the agent respond by a simple local rule, win-stay/lose-sample.
After a rewarding excursion the agent keeps the current affordance, and otherwise it draws a
fresh one from $\lambda$. This rule changes only how often each mode is started. Its steady
state is the tilted frequency $\lambda e^{\beta\Psi}$, where $\beta$ measures how strongly the
agent responds to reward [SM], so the occupation becomes $p_\beta\propto m\lambda
e^{\beta\Psi}$. Reward therefore acts on frequency as an external field, in the role a
magnetic field plays for a spin system, and leaves duration untouched. Because dominance is
measured in time, a reward that favors a short behavior must first overcome the advantage
that long behaviors gain from length-biased sampling. We call this advantage the
\emph{entropic force}, because it arises purely from how time is counted. The crossover
between the two regimes is abrupt. Two modes exchange dominance at the field
$\beta^\ast=[\log(m_1/m_2)+\log(\lambda_1/\lambda_2)]/(\Psi_2-\Psi_1)$, as in a first-order
phase transition, and the price of crossing is the dwell ratio $\log(m_1/m_2)$, the analogue
of a latent heat. In the box, reward for reaching the wall favors drift, yet the field must
reach $\beta\approx4$ before drift displaces circling [Fig.~\ref{fig:box}(c)].

\paragraph*{Equation of state.}
The geometry of the domain also has a conjugate.
Differentiating the free energy with respect to the volume of the domain gives the pressure,
which takes an ideal-gas form,
\begin{equation}
\Pi\,\vol=T\,\langle\zeta\rangle_{p_0},\qquad \zeta(a):=\frac{\partial\log m(a)}{\partial\log\vol},
\label{eq:eos}
\end{equation}
where $\zeta(a)$ measures how fast the dwell of mode $a$ grows as the domain is enlarged. For
a mode that escapes by diffusion the dwell grows as the square of the linear size, so
$\zeta=2/d$ in $d$ dimensions. For a mode that crosses ballistically it grows linearly, so
$\zeta=1/d$. The pressure therefore measures how much of the record is trapped and how much
is ballistic.

\paragraph*{Verification in three systems.}
We test each relation in three systems chosen to share as little mechanism as possible. In
each, the prediction uses only dwells measured without reward, and it is compared with an
independent simulation of the dynamics [SM]. In the box ($d=2$) the measured exponents are
$\zeta=0.964$ for circling and $0.496$ for drift, against the predicted $1$ and $1/2$. When
win-stay/lose-sample is simulated with a real reward field, circling and drift share the
record equally at $\beta=3.89\pm0.04$, against $3.85$ predicted from the unrewarded dwells.
The global free energy computed
from the mean cycle time agrees with the integral of the local pressure to within $1$ to
$3\%$. In a one-dimensional decision process $\zeta$ runs from $2/d=2$ to $1/d=1$, following
the prediction to within $0.016$. Finally, a Braitenberg vehicle~\cite{Braitenberg1984}, a
robot whose affordance is a sensorimotor \emph{wiring} that is rewired at each wall contact,
has an occupation that matches $m\lambda/Z_0$ to within $1.1\times10^{-3}$ in total
variation [SM].

\begin{figure}[t]
\centering
\includegraphics[width=\columnwidth]{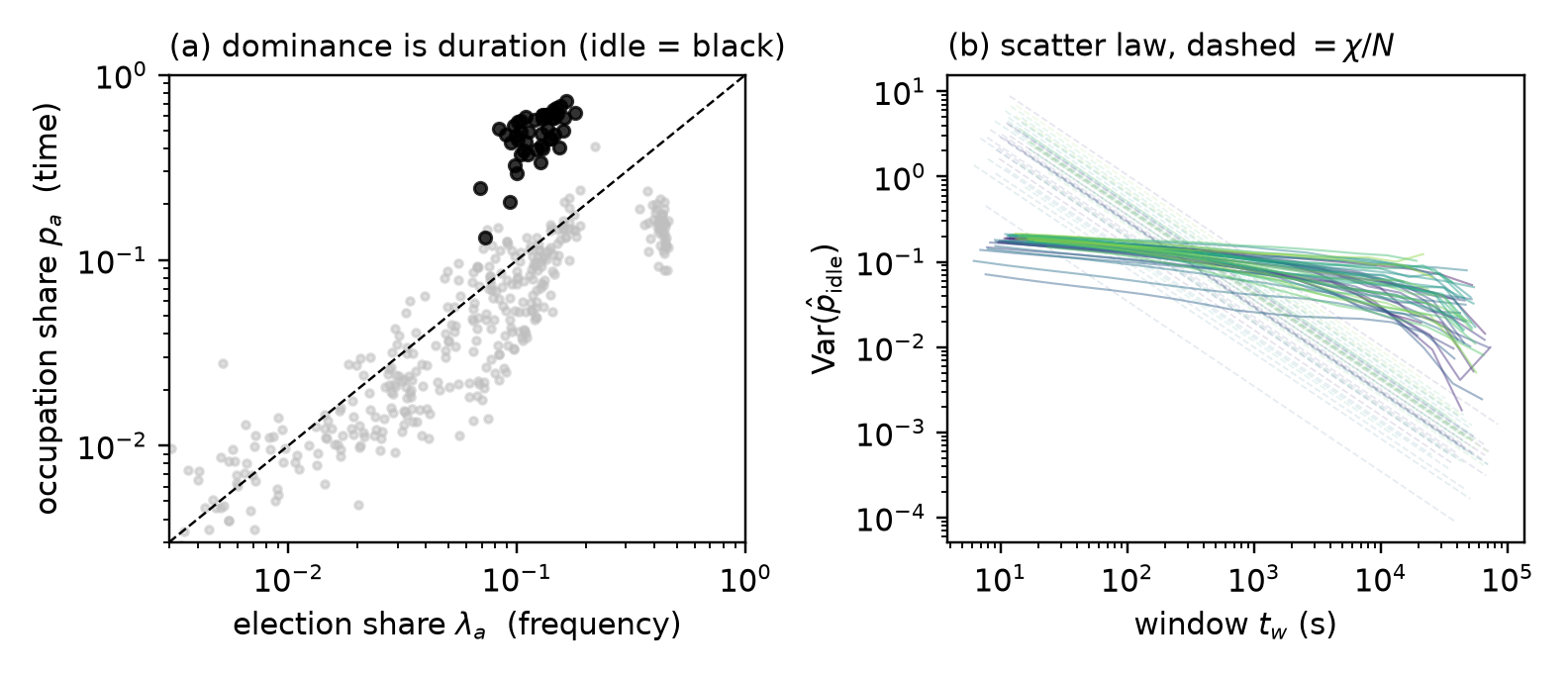}
\caption{The 47 fly records~\cite{McKenzieSmithData}. (a)~Election share (frequency) against
occupation share (time) for every fly and state, idle in black. (b)~Variance of the idle
occupation across windows of length $t_w$ against the finite-variance law $\chi/N$ of P2
(faint dashed, slope $-1$). Every fly decays more slowly.}
\label{fig:fly}
\end{figure}

\paragraph*{The fly records.}
We test the two variables on 47 multi-day recordings of
\emph{Drosophila}~\cite{McKenzieSmithData}, each $1.3$ to $7.8$ days long and labeled into
nine behavioral states [SM, Fig.~\ref{fig:fly}]. In the data an excursion is a bout. The
counterpart of the frequency $\lambda$ is the \emph{election share} of a state, the fraction
of bouts that begin in it, and the counterpart of the dwell $m$ is its mean bout duration.
The idle state is elected on a median $13\%$ of bouts yet fills $49\%$ of the record, and the
most-elected state differs from the most-occupied one in 46 of 47 flies
[Fig.~\ref{fig:fly}(a)]. Dominance is manufactured by duration.

Duration also carries individuality. When each record is split into two halves, the mean
idle dwell of a fly replicates between the halves at $r=0.85$, so that $91\%$ of its
variation across flies is a stable trait of the individual (split-half reliability
corrected to full record length). The election share replicates at only $r=0.24$, with
$39\%$ stable. The identity of a fly is carried by its duration scale, and its frequency is
the fluctuating part. Finally, the records show anomalous scatter. In
every fly the scatter of the idle occupation across windows decays with an exponent between
$-0.07$ and $-0.50$, where the finite-variance law gives $-1$ [Fig.~\ref{fig:fly}(b)], and
the idle durations have tail exponents from $0.56$ to $1.70$, median $0.77$. Both are
consistent with a repertoire near the ergodic edge, where this anomalous scatter is expected.

\paragraph*{Discussion.}
Several features of behavior that are usually read as signs of design turn out to be
consequences of one free energy. Dominance is the Gibbs weight, the matching law is
length-biased sampling, apparent optimality is the ground-state collapse of the third law,
and criticality is the ground state itself. All of them are properties of duration, the
variable that theories of choice set aside. Once this null part is subtracted, what remains
is what a genuine account of reward must explain, and each law comes with its own test
(Table~\ref{tab:laws} in the SM). The laws are verified in a diffusing particle, a decision
process, and a self-rewiring robot, and the fly records show the duration structure they
rest on. Like the laws of thermodynamics across substances, they are theorems of the renewal
structure, and so they carry over to any mechanism that realizes it.

\begin{acknowledgments}
D.Y.T. was supported by CNPq grants 421955/2023-6 and 445096/2024-1, FAPESP grant 2013/07699-0, and Serrapilheira grant R-2401-47364. A.E.H. acknowledges support by the DFG German Research Foundation (EXC 2117-422037984) and the Human Frontier Science Program (RGP006/2025). We thank the Princeton \emph{Drosophila} data initiative for the public dataset.
\end{acknowledgments}

\clearpage
\onecolumngrid
\setcounter{secnumdepth}{2}
\setcounter{section}{0}
\setcounter{equation}{0}
\setcounter{table}{0}
\setcounter{figure}{0}
\renewcommand{\thesection}{S\arabic{section}}
\renewcommand{\theequation}{S\arabic{equation}}
\renewcommand{\thetable}{S\arabic{table}}
\renewcommand{\thefigure}{S\arabic{figure}}

\begin{center}
{\large\bfseries Supplemental Material for\\[2pt] Statistical mechanics of spontaneous behavior}\\[6pt]
Daniel Yasumasa Takahashi and Ahmed El Hady
\end{center}
\medskip

This Supplemental Material gives the proofs of the results stated in the Letter, together with
the numerical methods behind every quoted number. Section and equation numbers prefixed by S
refer to this document, and unprefixed numbers refer to the Letter. Throughout we assume that
relaxation within a basin is fast compared with the dwell, so that each excursion is
quasi-stationary, that the embedded affordance chain is positive Harris recurrent, so that it has
a unique stationary law, and that the mean cycle time is finite, $Z_0<\infty$. The small-noise
results of Sec.~\ref{sm:dict} further assume that the noise $\sigma$ is small.
Section~\ref{sm:third} is the one exception, since there the failure of $Z_0<\infty$ is itself
the subject. Table~\ref{tab:laws} collects the four laws with their corollaries and tests.

\begin{table}[htb]
\caption{The four laws of the affordance-reset thermodynamics. Each law is a theorem of postulates P0 to P3. The table gives its statement, its corollaries, and the test that checks it.}
\label{tab:laws}
\begin{ruledtabular}
\begin{tabular}{p{0.09\columnwidth}p{0.42\columnwidth}p{0.34\columnwidth}}
Law & Statement & Corollaries / test\\
\colrule
0th & an intensive $\mathrm v_i$ exists, equalizes, is transitive & matching law
$s=T/T_\mu$ [Eqs.~\eqref{eq:alloc},\eqref{eq:sens}], no learning\\
1st & every change of $U$ is heat or work, one function $F$ behind every conjugate
[Eqs.~\eqref{eq:diff},\eqref{eq:first}] & Maxwell relations, global $=$ local
(non-circular), equation of state [Eq.~\eqref{eq:eos}]\\
2nd & $D(q_n\Vert\rho)\downarrow$ and $\dot S_{\rm obs}\ge0$ [Eq.~\eqref{eq:second}] & monotone
relaxation, arrow of time measurable in data\\
3rd & degeneracy of $p_0$ $=$ divergence of the clock $Z_0$ [Eq.~\eqref{eq:third}],
unattainable in finite time & ground-state collapse, critical point $=$ ground state, aging
$N\sim t_w^{\alpha}$ beyond the edge\\
\end{tabular}
\end{ruledtabular}
\end{table}

\section{The occupation law}
\label{sm:occ}

We prove Eq.~\eqref{eq:gibbs}. Between two consecutive contacts with the constraint, the
affordance stays fixed at a value drawn from $\lambda$ for a random duration $\tau$ with
$\E[\tau\mid a]=m(a)$. The sequence of pairs $(\xi_i,\tau_i)$, in which the $\xi_i\sim\lambda$ are
independent and identically distributed (i.i.d.) and $\tau_i$ is the $i$th dwell, is therefore a
\emph{marked renewal process}, a renewal process whose cycles each carry a label. For a bounded
test function $g$, the renewal reward theorem gives the long-run time average
\begin{equation}
\lim_{t\to\infty}\frac1t\int_0^t g(a_{t'})\,\rd t'
=\frac{\E[\,g(\xi)\,\tau\,]}{\E[\tau]}
=\frac{\int g(a)\,m(a)\,\lambda(\rd a)}{\int m\,\rd\lambda}\qquad\text{a.s.},
\label{eq:sm-renewal}
\end{equation}
because the numerator is the expected reward $\int_0^{\tau}g\,\rd t'=g(\xi)\tau$ accumulated per
cycle and the denominator is the mean cycle length. Identifying the right-hand side with
$\int g\,\rd p_0$ yields
\begin{equation}
p_0(\rd a)=\frac{m(a)\,\lambda(\rd a)}{Z_0},\qquad Z_0=\int m\,\rd\lambda=\langle\tau\rangle_\lambda ,
\label{eq:sm-gibbs}
\end{equation}
valid whenever $Z_0<\infty$. If the affordance is instead redrawn from a Markov kernel $K$, the
excursions form a Markov-renewal process, and the same length-biased argument applied to the
stationary law $\rho$ of the embedded chain $(\xi_i)$ gives
$p(\rd a)=m(a)\,\rho(\rd a)/\int m\,\rd\rho$~\cite{Cinlar1975}, of which
Eq.~\eqref{eq:sm-gibbs} is the special case $\rho=\lambda$. The condition $Z_0<\infty$ is
precisely what makes the free energy $F=-T\log(Z_0/\tau_0)$ well defined, and its failure, which
is the loss of ergodicity, is taken up in Sec.~\ref{sm:third}. Finally, the almost-sure
convergence in Eq.~\eqref{eq:sm-renewal} turns postulate P2, the equivalence of time and
ensemble averages, into a theorem.

\subsection{Finite records and the fluctuation law}
\label{sm:clt}
By P2 a finite record is a finite system, and its size is the number of completed excursions.
For a window $t_w=\sum_{k\le N}\tau_k$ and a measurable $A\subset\mathcal A$, the error of the
empirical occupation decomposes over excursions,
\begin{equation}
\hat p_A-p_A=\frac1{t_w}\sum_{k\le N}\eta_k,\qquad
\eta_k=\big(\mathds 1_A(\xi_k)-p_A\big)\tau_k ,
\label{eq:sm-decomp}
\end{equation}
where $\hat p_A$ is the empirical occupation measure and the $\eta_k$ are i.i.d.\ with mean zero,
the mean vanishing by Eq.~\eqref{eq:sm-gibbs} itself. The central limit theorem for renewal
reward processes~\cite{Feller1971} then gives
\begin{equation}
\Var(\hat p_A)=\frac{\chi_A}{N}+o(N^{-1}),\qquad
\chi_A=\frac{\E\big[(\mathds 1_A(\xi)-p_A)^2\,\tau^2\big]}{Z_0^{2}},\qquad N=\frac{t_w}{Z_0}.
\label{eq:sm-clt}
\end{equation}
In a numerical test on a five-mode ensemble with exponential dwells and $N=400$, the measured and
predicted variances over $4000$ independent records agree to $0.1\%$ (ratio $1.001$), with bias
$O(1/N)$. Equation~\eqref{eq:sm-clt} plays three roles. It is the finite-size scaling of the
theory, it predicts how the scatter of matching data shrinks with session length, and it serves
as the ruler for the third law (Sec.~\ref{sm:third}). Because a record resolves its own
occupation only to accuracy $\sqrt{\chi_AZ_0/t_w}$, any limit in which $Z_0$ diverges is one in
which the record can no longer display its equilibrium.

\section{The small-noise dictionary}
\label{sm:dict}

We justify Eq.~\eqref{eq:dict} by specializing the observable dynamics to a gradient diffusion
in a single basin,
\begin{equation}
\rd X_t=-\partial_x V(X_t,a)\,\rd t+\sigma\,\rd W_t ,
\label{eq:sm-grad}
\end{equation}
started at the bottom $x^\ast(a)$ of the well, with the constraint $C$ placed on the basin
boundary. The Freidlin and Wentzell theory of exit times~\cite{FreidlinWentzell2012} then gives
\begin{equation}
\lim_{\sigma\to0}\sigma^2\log\E^\sigma[\tau^a]
=\inf_{t>0,\,y\in C}\Big\{\tfrac12\!\int_0^t\!\big(\dot\phi_s+\partial_xV(\phi_s,a)\big)^2\rd s
:\phi_0=x^\ast,\ \phi_t=y\Big\}=:\bar J^a .
\label{eq:sm-ldp}
\end{equation}
For a gradient field the action-minimizing path is the time reversal of the noiseless relaxation
into the well, along which $\dot\phi=+\partial_xV$, so the integrand is $2\partial_xV\,\dot\phi$ and
\begin{equation}
\bar J^a=2\big[V(y^\ast(a),a)-V(x^\ast(a),a)\big],
\label{eq:sm-quasi}
\end{equation}
which is twice the height of the barrier to the easiest exit point $y^\ast$. The mean dwell
therefore grows as $\E^\sigma[\tau^a]\asymp e^{\bar J^a/\sigma^2}$, which is Kramers' law, and
writing $m(a)=e^{-E(a)/T}$ suggests the identification
\begin{equation}
T:=\sigma^2,\qquad E(a):=-\bar J^a ,
\end{equation}
to the leading exponential order. 

\section{Zeroth law, chemical potential and matching}
\label{sm:zeroth}

Partition the repertoire into disjoint sub-repertoires $A_i$ with propensities $\mu_i$
($\sum_i\mu_i=1$) and conditional laws $\lambda_i$, so $\lambda=\sum_i\mu_i\lambda_i$ and
\begin{equation}
Z_0=\sum_i\mu_iZ_i,\qquad Z_i:=\int_{A_i}m\,\rd\lambda_i ,
\end{equation}
with $Z_i$ the mean dwell conditioned on entering $A_i$. Allocating propensity among the groups
carries a mixing entropy at a temperature $T_\mu$, and the resulting free energy depends on
whether the sub-repertoires are pooled (annealed) or kept separate (quenched),
\begin{align}
\text{annealed:}\quad &F_{\rm a}(\mu)=-T\log\!\Big(\sum_i\mu_iZ_i\Big)+T_\mu\sum_i\mu_i\log\mu_i,\\
\text{quenched:}\quad &F_{\rm q}(\mu)=-T\sum_i\mu_i\log Z_i+T_\mu\sum_i\mu_i\log\mu_i .
\end{align}
The behavioral chemical potential is $\mathrm v_i=\partial F/\partial\mu_i$. Making $F$
stationary under the constraint $\sum_i\mu_i=1$ equalizes it across groups,
$\mathrm v_i=\mathrm v_j$, and because $\mathrm v_i$ is a single number this equality is
transitive, which is the content of the zeroth law. The equalized allocation is
\begin{equation}
\text{annealed:}\quad \mu_i\propto\exp\!\Big[\tfrac{T}{T_\mu}\tfrac{Z_i}{Z_0}\Big]
\quad\text{(self-consistent)},\qquad
\text{quenched:}\quad \boxed{\ \mu_i\propto Z_i^{\,s},\quad s=\frac{T}{T_\mu}=\frac{\sigma^2}{T_\mu}\ }
\label{eq:sm-matching}
\end{equation}
The quenched result is the generalized matching law with sensitivity $s$, so strict matching
corresponds to $T=T_\mu$ and undermatching ($s<1$) to $\sigma^2<T_\mu$. Time allocation is the
length-biased image of choice,
\begin{equation}
p_i=\frac{\mu_iZ_i}{\sum_j\mu_jZ_j},
\label{eq:sm-time}
\end{equation}
so uniform propensity $\mu_i=1/n$ already gives $p_i\propto Z_i$, which is strict matching in
time, with unit sensitivity and without any learning.

\section{First law, heat and work, and the state function}
\label{sm:first}

In this thermodynamics the first law states that a single state function stands behind every
operationally measured force, and that every change of the mean energy is exactly heat or work.
We first assemble the potential, then prove the decomposition, and finally verify exactness.

\subsection{The potential and its conjugates}
The per-excursion free energy is $F=-T\log(Z_0/\tau_0)$ and the total potential over a record of
$N$ excursions is $\Phi=NF$. From $p_0=e^{-E/T}\lambda/Z_0$,
\begin{equation}
S=-D(p_0\Vert\lambda)=\frac{\langle E\rangle_{p_0}}{T}+\log\frac{Z_0}{\tau_0}
=-\frac{\partial F}{\partial T},
\label{eq:sm-entropy}
\end{equation}
so the Gibbs entropy is the conjugate of temperature. Differentiating $F$ at fixed occupation
gives the remaining conjugates, namely the pressure $\Pi=-\partial_\vol F$, the mean reward
$\Upsilon=-\partial_\beta F=T\langle\Psi\rangle$, and the behavioral chemical potentials
$\mathrm v_i=\partial F/\partial\mu_i$ of Sec.~\ref{sm:zeroth}. Collecting them,
\begin{equation}
\rd F=-S\,\rd T-\Pi\,\rd\vol-\Upsilon\,\rd\beta+\sum_i\mathrm v_i\,\rd\mu_i ,
\label{eq:sm-diff}
\end{equation}
and for the whole record $\rd\Phi=N\,\rd F+F\,\rd N$, so that $F=\partial\Phi/\partial N$ is the
conjugate of the record size. Unlike its classical counterpart, this differential carries no
conserved quantity, because the energy $E=-\bar J^a$ is set by the environment, is not additive
over the repertoire, and is exchanged with no bath. What survives of the classical first law is
the decomposition of every change of $U$ into heat and work, and that decomposition survives
exactly.

\subsection{The heat and work decomposition}
\label{sm:heatwork}
We adopt the exact assignment $E(a):=-T\log\big(m(a)/\tau_0\big)$, whose small-noise limit is
Eq.~\eqref{eq:dict}. With it, $p_0=\lambda e^{-E/T}/(Z_0/\tau_0)$ holds at any noise level and the
identities below are exact. Let $U=\langle E\rangle_{p_0}$. Any smooth change of the controls at
fixed frequency gives $\rd U=\int E\,\rd p_0+\int p_0\,\rd E$, and the first term is exactly
$T\rd S$, because $\int\rd p_0=0$ implies
\begin{equation}
\rd S=-\!\int\rd p_0\,\log\frac{p_0}{\lambda}
=-\!\int\rd p_0\Big[-\frac{E}{T}-\log\frac{Z_0}{\tau_0}\Big]
=\frac1T\!\int E\,\rd p_0 .
\label{eq:sm-heat}
\end{equation}
The second term is the work done by the geometry. Section~\ref{sm:eos} shows that
$\int p_0\,\partial_\vol E\,\rd\vol=-\Pi\,\rd\vol$, so the two pressure meters
$-\partial_\vol F$ and $-\langle\partial_\vol E\rangle$ necessarily agree, and along quasistatic
paths at fixed frequency and field
\begin{equation}
\rd U=T\rd S-\Pi\,\rd\vol .
\label{eq:sm-firstlaw}
\end{equation}
Here heat is the reshuffling of the record over its modes, and work is the displacement of the
energies by the environment. A temperature path at fixed geometry is therefore pure heat
whenever the barriers do not depend on $T$, as in the Arrhenius regime of Sec.~\ref{sm:dict}.
On the discrete ensemble of the next subsection the stepwise identity $\int E\,\rd p_0=T\rd S$
holds to $2\times10^{-15}$. Integrated along a temperature path, $\Delta U=\int T\rd S$ holds to
$10^{-6}$, and along a volume path $\Delta U=T\Delta S-\int\Pi\,\rd\vol$ holds to the same
accuracy, with the two pressure meters coinciding to $10^{-6}$.

\subsection{Exactness and the Maxwell relations}
Because every term in Eq.~\eqref{eq:sm-diff} is a first partial derivative of the single function
$F$, the form is closed and its mixed second derivatives commute, giving
\begin{equation}
\frac{\partial\Pi}{\partial\beta}=\frac{\partial\Upsilon}{\partial\vol},\qquad
\frac{\partial S}{\partial\vol}=\frac{\partial\Pi}{\partial T},\qquad
\frac{\partial S}{\partial\beta}=\frac{\partial\Upsilon}{\partial T}.
\label{eq:sm-maxwell}
\end{equation}
We verify these on a solvable discrete model with $n$ affordances, each assigned a barrier
$E_a$, a frequency $\lambda_a$, a dwell-scaling exponent $\zeta_a$ (so that $\log m_a$ is affine
in $\log\vol$) and a reward $\Psi_a$, for which $F(T,\vol,\beta,\mu)$ is a closed-form
log-sum-exp. Each conjugate computed analytically agrees to machine precision with a centered
finite difference of $F$, the three Maxwell relations~\eqref{eq:sm-maxwell} hold to
$\sim\!10^{-11}$, and $\oint\rd F=0$ holds around closed loops in the $(T,\beta)$ plane to
$\sim\!10^{-12}$. Exactness fails by construction when the reward $\Psi$ itself depends on
$\vol$, because then $-\partial_\vol F=\Pi+T\beta\langle\partial_\vol\Psi\rangle_{p_\beta}\neq\Pi$.
A pressure that is a state function therefore requires a reward schedule that does not depend on
the confining geometry.

\section{Second law, relaxation and the arrow of the record}
\label{sm:second}

\subsection{The $H$-theorem}
Read the record as a sequence of excursions and keep only which mode each excursion used. This
sequence of modes is a Markov chain, the \emph{embedded chain}, whose transition rule
$K(a,\cdot)$ gives the probability of each next mode when the current mode is $a$, and whose
stationary law $\rho$ is the distribution of modes that the rule leaves unchanged, $\rho K=\rho$.
Without reward every mode is drawn afresh from the frequency $\lambda$, so $K(a,\cdot)=\lambda$
and $\rho=\lambda$. Under the win-stay/lose-sample rule of Sec.~\ref{sm:reward} the system
instead keeps its current mode with probability $\bar w(a)$ and otherwise draws a new one from
$\lambda$, so that $K(a,\cdot)=\bar w(a)\,\delta_a+(1-\bar w(a))\,\lambda$, where $\delta_a$ puts
all the weight on $a$, and $\rho\propto\lambda/(1-\bar w)$.

Let $q_n$ be the distribution of the mode at the $n$th excursion, so that $q_{n+1}=q_nK$. Its
distance from equilibrium is measured by the relative entropy
$D(q\Vert\rho)=\sum_a q(a)\log[q(a)/\rho(a)]$, which vanishes only when $q=\rho$. Passing two
distributions through the same random rule can only make them harder to tell apart, a fact known
as the data-processing inequality~\cite{CoverThomas2006}, $D(qK\Vert\rho K)\le D(q\Vert\rho)$,
and since $\rho K=\rho$ it gives
\begin{equation}
D(q_{n+1}\Vert\rho)\ \le\ D(q_n\Vert\rho).
\label{eq:sm-H}
\end{equation}
The distance to equilibrium therefore never grows, and it strictly shrinks whenever the rule
mixes, that is, whenever the chain can move between modes. Because the time occupation is the
distribution of modes weighted by their dwells (Sec.~\ref{sm:occ}), it inherits the same
descent. For a six-mode win-stay/lose-sample rule started from each of $20$ random
distributions, $D(q_n\Vert\rho)$ never increases by more than $10^{-13}$ per step, and the
stated $\rho$ is stationary to machine precision.

\subsection{The arrow of the record}
Let $\mathsf P$ be the probability of an observed sequence of behavioral states over a window,
and $\widetilde{\mathsf P}$ the probability of the same sequence played backward. The
irreversibility
\begin{equation}
\mathcal I=D\big(\mathsf P\,\Vert\,\widetilde{\mathsf P}\big)\ \geq\ 0
\label{eq:sm-I}
\end{equation}
is a relative entropy, so it is never negative, and it is zero only when the record looks the
same forward and backward~\cite{Seifert2012,Crooks1999}. The reset breaks this symmetry, because
at each contact many different states are sent to the same injection point and this map cannot
be run backward. The full path therefore produces entropy at the rate
$\dot S_{\rm obs}=\langle\dot S_{\rm exc}\rangle_{p_0}>0$ whenever resets occur, even when the
affordance is resampled symmetrically. Grouping states into behaviors can hide part of this
asymmetry but cannot create any, so the irreversibility $\mathcal I$ of the observed record is a
lower bound on $\dot S_{\rm obs}$.

\section{Third law and unattainability}
\label{sm:third}

\subsection{The dwell floor and the clock bound}
Every excursion must cross the domain, so at fixed geometry the mean dwell is bounded below by
the time of the fastest, ballistic crossing, $m(a)\ge\tau_{\min}>0$. For any measurable
$\mathcal G\subset\mathcal A$ this gives
\begin{equation}
p_0(\mathcal G^c)=\frac{\int_{\mathcal G^c}m\,\rd\lambda}{Z_0}
\ \ge\ \frac{\tau_{\min}\,\lambda(\mathcal G^c)}{Z_0}
\qquad\Longleftrightarrow\qquad
Z_0\ \ge\ \frac{\tau_{\min}\,\lambda(\mathcal G^c)}{p_0(\mathcal G^c)} .
\label{eq:sm-floor}
\end{equation}
Suppose now that the controls change so that the occupation concentrates on $\mathcal G$,
$p_0(\mathcal G^c)\to0$, while the modes outside $\mathcal G$ keep a frequency
$\lambda(\mathcal G^c)\ge\lambda_\ast>0$. The bound then forces the partition function to
diverge. Since $Z_0$ is the mean cycle time, the elementary renewal theorem~\cite{Feller1971}
gives an expected number of completed excursions in a window $t_w$ of
$\E N(t_w)\sim t_w/Z_0\to0$, and by the fluctuation law~\eqref{eq:sm-clt} the occupation of the
window keeps a variance $\chi/N$. No finite record therefore equilibrates at a degeneracy of the
occupation measure. Classical thermodynamics treats unattainability as an independent principle
or derives it from added resource assumptions~\cite{MasanesOppenheim2017}, whereas here it is a
theorem, because the normalizer of the equilibrium measure and the clock that samples it are the
same number.

\subsection{Ground-state collapse}
\label{sm:nernst}
As $T=\sigma^2\to0$, the integral $Z_0=\int e^{-E/T}\rd\lambda$ is dominated by the minimizers
of $E$ (Laplace's method), so
\begin{equation}
\lim_{T\to0}F=\min_a E(a)=-\sup_a\bar J^a,\qquad p_0\ \Longrightarrow\ \delta_{\mathcal G},\quad
\mathcal G=\arg\min_a E(a),
\label{eq:sm-third}
\end{equation}
which is the mode with the deepest barrier and hence the longest dwell. The entropy tends to
$S\to\log\lambda(\mathcal G)$, the log frequency of the ground-state set, so strict Nernst
behavior $S\to0$ would hold only if $\lambda(\mathcal G)=1$. This is the familiar $T\to0$
concentration of any Gibbs measure, and it is what apparent optimality and selection reduce to.
The preceding subsection shows, however, that no finite record attains this limit. Displaying
it would take a recording time $t_w\gg Z_0\ge\lambda(\mathcal G)\,m(\mathcal G)
=\lambda(\mathcal G)\,e^{\bar J_{\mathcal G}/T}$, which diverges in Arrhenius fashion, and at
$T=0$ the renewal process has no renewals at all, so the whole record is a single unfinished
excursion.

\subsection{The ergodic edge and critical slowing down}
\label{sm:edge}
Consider the trap ensemble~\cite{Bouchaud1992,MonthusBouchaud1996}, in which barriers on
$[0,\infty)$ have the exponential frequency $\lambda(\bar J)=T_\lambda^{-1}e^{-\bar J/T_\lambda}$,
decaying with depth at rate $1/T_\lambda$, and dwells $m=e^{\bar J/T}$. Its partition function is
elementary,
\begin{equation}
Z_0=\int_0^\infty e^{\bar J\left(\frac1T-\frac1{T_\lambda}\right)}\frac{\rd\bar J}{T_\lambda}
=\frac{T}{T-T_\lambda},\qquad T>T_\lambda ,
\label{eq:sm-trapZ}
\end{equation}
and it diverges as $(T-T_\lambda)^{-1}$ at the edge $T=T_\lambda$. In ordinary critical
phenomena the partition function stays finite while correlations diverge. Here the divergence of
$Z_0$ is itself the critical slowing down, so thermodynamics and kinetics meet in a single
quantity, and by the bound of Sec.~\ref{sm:third} no finite record can display the edge. The
dwells have the power-law tail $\mathbb P(\tau>x)=x^{-\alpha}$ with $\alpha=T/T_\lambda$. For
$\alpha>1$ renewals accumulate at the steady rate $1/Z_0$, whereas for $\alpha<1$ the mean dwell
is infinite, sums of dwells are governed by a one-sided stable law, and the number of renewals
grows only as $N(t_w)\sim t_w^{\alpha}$, up to slowly varying corrections~\cite{Feller1971}. %

\subsection{The critical point is the ground state}
\label{sm:soc}
Because the Boltzmann weight is the dwell, the occupation concentrates wherever the dwell is
longest. Near a continuous transition the passage time diverges, which is critical slowing down,
so the longest-dwelling affordance is the critical point and $E=-T\log m$ has its minimum there.
We make this concrete with the saddle-node bottleneck of Fig.~\ref{fig:soc}, the canonical form
of critical slowing down and the channel of type-I intermittency, in which the affordance is the
control parameter $\epsilon$ of
\begin{equation}
\rd x=(\epsilon+x^2)\,\rd t+\sigma\,\rd W_t
\label{eq:sm-saddle}
\end{equation}
on $[-X_0,X_0]$. The trajectory is reinjected at $-X_0$ on reaching $+X_0$, which is the
constraint, and $\epsilon$ is redrawn at every contact from the uniform frequency on $(0,1]$, so
that $\epsilon_c=0$ is the critical point. The deterministic passage time is exact,
\begin{equation}
m(\epsilon)=\int_{-X_0}^{X_0}\frac{\rd x}{\epsilon+x^2}
=\frac{2}{\sqrt\epsilon}\arctan\frac{X_0}{\sqrt\epsilon}\ \xrightarrow[\epsilon\to0^+]{}\ \frac{\pi}{\sqrt\epsilon},
\label{eq:sm-mcrit}
\end{equation}
and is maximal at $\epsilon_c$. In simulation the dwell matches Eq.~\eqref{eq:sm-mcrit} to better
than $0.1\%$ for $\epsilon>0.3$ at $\sigma=0.08$. The entropic force
$\partial_\epsilon\log m\to-1/2\epsilon$ points inward and diverges at criticality. This holds
near any continuous transition, since a passage time $m(a)\sim|a-a_c|^{-\gamma}$ produces a force
$\sim-\gamma/(a-a_c)$ that always points toward $a_c$. The peak does require a marginal-passage
geometry, however. In a simple confining well $m$ varies monotonically with stiffness and has no
interior maximum, which is why the relevant critical structure is the bottleneck.

By Eq.~\eqref{eq:sm-gibbs} the occupation is
$p(\epsilon)\propto m(\epsilon)\lambda(\epsilon)\sim\epsilon^{-1/2}$ as $\epsilon\to0^+$. This
square-root singularity is integrable, so the occupation concentrates at criticality while
remaining normalizable, even though the mean dwell
$\int m\,p\,\rd\epsilon\sim\int\epsilon^{-1}\rd\epsilon$ diverges logarithmically. In simulation
at $\sigma=0.15$, the time occupation generated from the uniform frequency is tilted $5.8$-fold
toward $\epsilon_c$ across $(0,1]$ [Fig.~\ref{fig:soc}(a)], and it follows the noiseless curve
$m\lambda/Z_0$ everywhere except in the first bin, where noise shortens the longest passages.

The durations are scale-free. Changing variables to $\tau=m(\epsilon)$ with uniform $\lambda$
gives $P(\tau)=\lambda(\epsilon)\,|\rd\epsilon/\rd\tau|$, which for $m\approx\pi\epsilon^{-1/2}$
equals $2\pi^2\tau^{-3}$. Noise cuts this power law off, and balancing drift against noise near
the bottleneck, where $x\sim\epsilon^{1/2}$, gives a crossover $\epsilon^\ast\sim\sigma^{4/3}$ and
a duration cutoff $\tau^\ast\sim\sigma^{-2/3}$. The power law therefore holds for
$2m(\epsilon{=}1)<\tau<\tau^\ast$ and extends further as the noise falls. For every $\sigma$ from
$0.4$ to $0.08$ the simulated $P(\tau)$ follows the noiseless law up to the cutoff, with a tail
exponent of $3.2\pm0.1$ fitted between $2m(\epsilon=1)$ and the cutoff ($3.04$ at the lowest
noise), and the curves collapse under $P(\tau)=\tau^{-3}\mathcal F(\tau\sigma^{2/3})$
[Fig.~\ref{fig:soc}(b), inset]. The exponent is not universal, however. Near a general
transition with passage divergence $m\sim|a-a_c|^{-\gamma}$ and frequency
$\lambda\sim|a-a_c|^{\kappa}$, the same change of variables gives
$P(\tau)\sim\tau^{-1-(1+\kappa)/\gamma}$, which recovers $3$ for the saddle-node with uniform
frequency ($\gamma=\tfrac12$, $\kappa=0$). Shallower divergences, or frequencies that thin out
toward the transition, lower the exponent, so a measured duration exponent reflects the geometry
of the bottleneck and the frequency jointly.

Observed criticality is usually taken as evidence of tuning, selection, or a self-organizing
feedback~\cite{BakTangWiesenfeld1987,Munoz2018}. Here, by contrast, the system sits at a
critical point simply because the critical point is the longest-dwelling state and length-biased
sampling converts long dwell into high occupation. No parameter is tuned to $\epsilon_c$ and
nothing drives the control parameter there, since it is drawn afresh from a fixed frequency at
every contact and only the time occupation is reweighted. This dwell-weighted proximity to
criticality differs from avalanche self-organized criticality, in which the critical state is an
attractor of the dynamics. Criticality is therefore a null expectation that must be excluded
before an observed critical point is attributed to adaptation, just as the neutral occupation is
the null for dominance.

\subsection{A frozen record, numerically}
\label{sm:frozen}
We simulate three modes with barriers $\bar J=(0,\,0.8,\,1.0)$, frequency
$\lambda=(0.8,\,0.1,\,0.1)$ and exponential dwells, using a fixed window $t_w=10^4$ and $300$
records per temperature. In equilibrium the shallower of the two trapped modes holds a share
$f_0=\big[1+e^{0.2/T}\big]^{-1}\to0$ of the time spent in the traps, but a finite record freezes
into whichever trap it enters first.
\begin{center}
\begin{tabular}{ccccc}
$T$ & $Z_0$ & $\E N(t_w)$ & $f_0$ (equilibrium) & $\hat f$ (measured) \\
\hline
$0.60$ & $1.7$ & $5852$ & $0.417$ & $0.417\pm0.019$ \\
$0.40$ & $2.8$ & $3627$ & $0.378$ & $0.376\pm0.024$ \\
$0.25$ & $8.7$ & $1148$ & $0.310$ & $0.308\pm0.042$ \\
$0.15$ & $100$ & $100$ & $0.209$ & $0.225\pm0.106$ \\
$0.10$ & $2.5\times10^3$ & $4.0$ & $0.119$ & $0.288\pm0.358$ \\
$0.08$ & $2.9\times10^4$ & $0.34$ & $0.076$ & $0.452\pm0.478$ \\
\end{tabular}
\end{center}
Down to $N\approx10^2$ renewals the record tracks equilibrium with the $\sqrt{\chi/N}$ scatter of
Eq.~\eqref{eq:sm-clt}, but at $N=4$ it no longer does, and at $N<1$ the window is a single
unfinished excursion whose occupation is close to a coin flip between the traps. The collapse of
Sec.~\ref{sm:nernst} is thus real yet unattainable, since it is approached only on records whose
length grows with the clock.

\section{Equation of state}
\label{sm:eos}

From $F=-T\log Z_0$ and $\partial_\vol\log Z_0=\langle\partial_\vol\log m\rangle_{p_0}$,
\begin{equation}
\Pi=-\frac{\partial F}{\partial\vol}=T\big\langle\partial_\vol\log m\big\rangle_{p_0}
\ \Longleftrightarrow\ \Pi\vol=T\langle\zeta\rangle_{p_0},\qquad
\zeta(a)=\frac{\partial\log m(a)}{\partial\log\vol}.
\label{eq:sm-eos}
\end{equation}
Under isotropic dilation of a fixed shape, a mode that escapes by diffusion has
$m\sim\ell^2/2\sigma^2$ and hence $\zeta=2/d$, while a ballistic mode has $m\sim\ell/|\mathbf v|$
and hence $\zeta=1/d$. For the decision process with absorbing thresholds at $\pm c$ and drift
$\nu$ the mean dwell is known in closed form~\cite{Bogacz2006}, $m(c)=(c/\nu)\tanh(\nu c/\sigma^2)$,
and with $u=\nu c/\sigma^2$ it gives
\begin{equation}
\zeta(c)=1+\frac{2u}{\sinh 2u},
\label{eq:sm-zetac}
\end{equation}
which runs from $\zeta\to2$ as $u\to0$, where escape is limited by diffusion, to $\zeta\to1$ as
$u\to\infty$, where it is limited by drift. Taking $\vol=c$, the one-dimensional equation of
state reads $\Pi c=T\langle\zeta\rangle$, which approaches the ideal-gas form $\Pi c=T=\sigma^2$
at large threshold.

\section{Reward field and coexistence}
\label{sm:reward}

Under the win-stay/lose-sample rule, the affordance of each excursion is retained with a
probability $w(a,\varrho)$ that depends on the reward signal $\varrho$ the excursion produced, and
is otherwise redrawn from $\lambda$. The embedded chain then has kernel
$K(a,\cdot)=\bar w(a)\,\delta_a+(1-\bar w(a))\,\lambda$ with $\bar w(a)=\E[w\mid a]$, and its unique
stationary law is
\begin{equation}
\rho(\rd a)\propto\frac{\lambda(\rd a)}{1-\bar w(a)} .
\label{eq:sm-mech}
\end{equation}
When the mean retention is exponential, $1-\bar w(a)=e^{-\beta\Psi(a)}$, this gives
$\rho\propto\lambda e^{\beta\Psi}$ and hence, by Sec.~\ref{sm:occ},
$p_\beta\propto m\,\lambda\,e^{\beta\Psi}$, so the reward tilt is the exact stationary law of a
local rule driven by experience. Two modes then carry equal weight when
$m_1\lambda_1e^{\beta\Psi_1}=m_2\lambda_2e^{\beta\Psi_2}$, which defines the first-order
coexistence line
\begin{equation}
\beta^\ast=\frac{\log(m_1/m_2)+\log(\lambda_1/\lambda_2)}{\Psi_2-\Psi_1},
\label{eq:sm-coex}
\end{equation}
along which the quantity exchanged between the modes, the analogue of a latent heat, is the log
dwell ratio $\log(m_1/m_2)$.

\section{The fly records}
\label{sm:fly}

\begin{figure*}[t]
\centering
\includegraphics[width=0.98\textwidth]{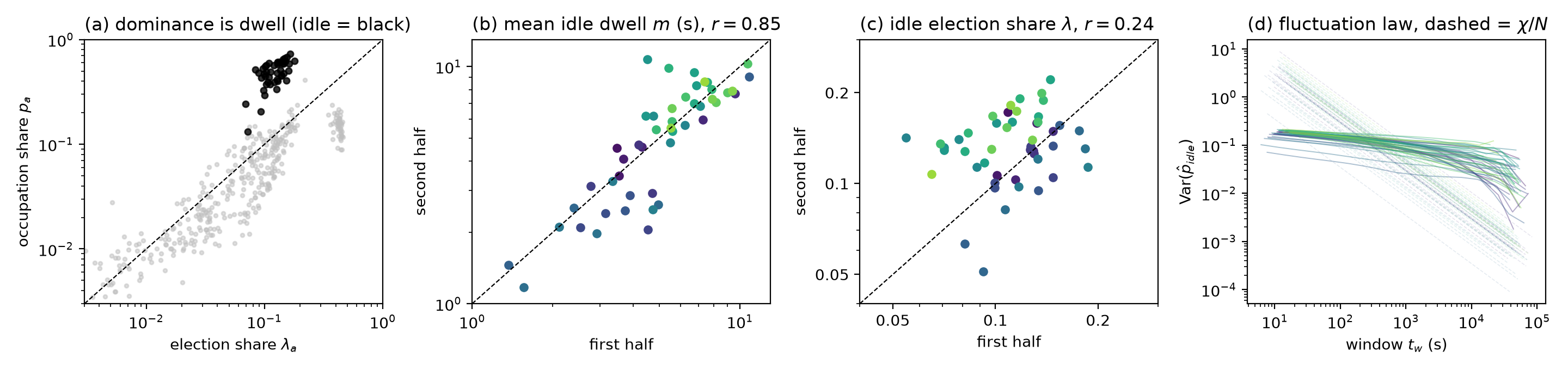}
\caption{The 47 multi-day fly records~\cite{McKenzieSmithData}. (a) Election share
against occupation share for every fly and state, idle in black. Idle sits far above the
diagonal and the most-entered state, the grey cluster near $\lambda=0.4$, sits below it.
Dominance is manufactured by dwell. (b) The mean idle dwell of each fly, first half of
the record against second half. The dwell scale replicates at $r=0.85$ and identifies the
individual. (c) The same split for the idle election share, which replicates at $r=0.24$.
(d) Variance of the idle occupation across non-overlapping windows of length $t_w$ (solid,
one curve per fly) against the parameter-free finite-variance law $\chi/N$ (faint dashed).
Every fly decays far more slowly than the predicted slope $-1$, the anomalous scatter of
a repertoire near the ergodic edge.}
\label{fig:sm-fly}
\end{figure*}

The analysis includes all 47 flies of the Princeton long-timescale
dataset~\cite{McKenzieSmithData}, recorded in four cohorts during 2022. Each fly contributes 1.3
to 7.8 days of continuous ethogram over nine behavioral states, amounting to
$0.7\times10^5$ to $6.8\times10^5$ completed bouts per record. Bout lists of states and
durations are read off the published per-frame labels without smoothing. Because the idle state
holds the largest time share in 45 of the 47 flies, it serves as the reference mode below.
Table~\ref{tab:sm-fly} lists the per-fly numbers and Fig.~\ref{fig:sm-fly} shows the
corresponding panels, of which (a) and (d) appear as Fig.~\ref{fig:fly} of the Letter.

\emph{Dominance is manufactured by dwell.} The occupation factorizes exactly as
$p_a=\lambda_a m_a/Z_0$, where $\lambda_a$ is the election share, the fraction of completed bouts
that enter $a$, and $m_a$ is the mean dwell. If dominance reflected preference, the structure
would sit in $\lambda$, but the data place it in $m$. The flies elect idle on a median $0.13$ of
bouts (range $0.07$ to $0.18$) yet spend a median $0.49$ of the record in it (range $0.13$ to
$0.73$), a four to six fold length bias. The most-entered state is the unstereotyped state in
all 47 flies, whereas the most-occupied state is idle in 45, and the two coincide in only a
single fly [Fig.~\ref{fig:sm-fly}(a)]. Across the full nine-state repertoire the variance of log
occupation is carried mainly by the election shares, with the dwell accounting for a median
variance fraction of $0.18$ per fly. Election thus decides which states appear in the repertoire
at all, while dwell decides which of them dominates the record.

\emph{Individuality lives on the dwell axis.} Across flies the variance of
$\log p_{\rm idle}=\log\lambda+\log m-\log Z_0$ splits into a dwell fraction of $1.39$ and an
election fraction of $0.44$. Fractions can exceed one because the covariance terms with
$\log Z_0$ are negative and absorb the balance. Consistently, $\log p_{\rm idle}$ correlates with
$\log m$ at $0.90$ but with $\log\lambda$ only at $0.65$. To separate a stable trait from
finite-record noise we split every record into two halves by time. The mean idle dwell
replicates between the halves at $r=0.85$ (Spearman $0.82$), so that $91\%$ of its across-fly
variance is stable after the Spearman-Brown correction to full record length, and the occupation
follows at $r=0.72$. The election share, in contrast, replicates at only $r=0.24$, with $39\%$
stable [Fig.~\ref{fig:sm-fly}(b,c)]. The identifying trait of an individual fly is therefore its
dwell scale, while its election rate is the fluctuating component. Two confounds can be set
aside. The correlation of $\log m$ with record length is only $0.17$, so the length bias of
heavy-tailed sample means contributes little. The two halves do share arena, camera, and week,
which could inflate all replication numbers together, but a shared setting cannot produce the
contrast between $m$ and $\lambda$.

\emph{The finite-variance scatter law fails in every fly.} Postulate P2 fixes the scatter that a
finite record can have, $\Var(\hat p)=\chi/N$ with the plug-in $\chi$ of Eq.~\eqref{eq:sm-clt}.
In every one of the 47 flies, however, the measured variance of idle occupation across
non-overlapping windows decays far more slowly, with log-log slopes from $-0.07$ to $-0.50$
against the predicted $-1$ [Fig.~\ref{fig:sm-fly}(d)], as expected in the anomalous regime of a
repertoire near the ergodic edge (Sec.~\ref{sm:edge}). The idle dwell tail has survival exponents
near unity, $0.56$ to $1.70$ by the maximum-likelihood fit of Ref.~\cite{Clauset2009} and $0.75$
to $1.84$ by the Hill estimator~\cite{Hill1975}. Single records in this regime have random growth
exponents, and the bout count averaged over flies grows sublinearly, as $t_w^{0.93}$ over three
decades. As a check, the analysis pipeline recovers ground truth on seeded Pareto ensembles,
returning growth exponents of $1.00$, $1.00$, $0.67$ and $0.50$ against $\min(1,\alpha)$ for an
exponential control and for $\alpha=1.5$, $0.7$ and $0.5$.

\begin{table*}[p]
\caption{The 47 flies of Ref.~\cite{McKenzieSmithData}. Columns give the fly identifier,
the 2022 recording date and camera, the record length in days, the number of completed
bouts, the mean bout time $Z_0$, the idle occupation share $p_{\rm idle}$, the idle
election share $\lambda$, the mean idle dwell $m$, and the measured decay exponent of the
idle occupation scatter across windows, whose finite-variance value would be $-1$.}
\label{tab:sm-fly}
\small
\renewcommand{\arraystretch}{0.95}
\begin{ruledtabular}
\begin{tabular}{cccrrrrrrr}
fly & date & cam & days & bouts & $Z_0$\,(s) & $p_{\rm idle}$ & $\lambda$ & $m$\,(s) & Var \\
\hline
1 & 02-17 & 1 & 4.9 & 391\,286 & 1.08 & 0.51 & 0.14 & 4.05 & $-0.13$ \\
2 & 02-17 & 1 & 3.3 & 295\,147 & 0.98 & 0.37 & 0.10 & 3.50 & $-0.26$ \\
3 & 02-17 & 1 & 6.2 & 450\,054 & 1.18 & 0.48 & 0.15 & 3.89 & $-0.23$ \\
4 & 02-17 & 1 & 5.0 & 388\,780 & 1.11 & 0.43 & 0.11 & 4.44 & $-0.32$ \\
5 & 02-17 & 2 & 4.0 & 238\,256 & 1.47 & 0.58 & 0.13 & 6.58 & $-0.44$ \\
6 & 02-17 & 2 & 6.8 & 305\,636 & 1.92 & 0.66 & 0.15 & 8.53 & $-0.40$ \\
7 & 02-17 & 2 & 7.8 & 568\,555 & 1.18 & 0.48 & 0.13 & 4.44 & $-0.26$ \\
8 & 02-17 & 2 & 6.1 & 466\,330 & 1.13 & 0.41 & 0.13 & 3.67 & $-0.26$ \\
9 & 02-17 & 3 & 6.3 & 566\,283 & 0.96 & 0.40 & 0.13 & 2.96 & $-0.31$ \\
10 & 02-17 & 3 & 4.3 & 370\,537 & 0.99 & 0.39 & 0.12 & 3.21 & $-0.19$ \\
11 & 02-17 & 3 & 3.8 & 412\,364 & 0.79 & 0.29 & 0.10 & 2.30 & $-0.23$ \\
12 & 02-17 & 3 & 6.5 & 680\,382 & 0.83 & 0.33 & 0.10 & 2.76 & $-0.29$ \\
13 & 02-17 & 4 & 4.2 & 353\,781 & 1.03 & 0.45 & 0.14 & 3.34 & $-0.23$ \\
14 & 02-17 & 4 & 3.1 & 292\,561 & 0.92 & 0.37 & 0.11 & 3.09 & $-0.29$ \\
15 & 02-17 & 4 & 5.8 & 222\,122 & 2.27 & 0.68 & 0.16 & 9.90 & $-0.50$ \\
16 & 03-13 & 3 & 3.5 & 399\,571 & 0.77 & 0.13 & 0.07 & 1.40 & $-0.21$ \\
17 & 03-13 & 3 & 3.9 & 486\,935 & 0.69 & 0.24 & 0.07 & 2.45 & $-0.19$ \\
18 & 03-13 & 3 & 4.8 & 364\,258 & 1.15 & 0.50 & 0.16 & 3.60 & $-0.18$ \\
19 & 03-13 & 3 & 4.7 & 658\,925 & 0.62 & 0.21 & 0.09 & 1.37 & $-0.21$ \\
20 & 03-13 & 4 & 4.3 & 409\,322 & 0.91 & 0.41 & 0.15 & 2.44 & $-0.19$ \\
21 & 03-13 & 4 & 3.2 & 352\,333 & 0.80 & 0.34 & 0.13 & 2.11 & $-0.23$ \\
22 & 03-13 & 4 & 4.4 & 421\,395 & 0.91 & 0.39 & 0.11 & 3.31 & $-0.09$ \\
23 & 03-13 & 4 & 2.3 & 183\,990 & 1.10 & 0.45 & 0.14 & 3.56 & $-0.15$ \\
24 & 03-26 & 3 & 3.8 & 299\,107 & 1.10 & 0.48 & 0.09 & 5.86 & $-0.15$ \\
25 & 03-26 & 3 & 3.0 & 230\,023 & 1.14 & 0.45 & 0.10 & 5.09 & $-0.12$ \\
26 & 03-26 & 3 & 4.8 & 354\,918 & 1.16 & 0.49 & 0.10 & 5.54 & $-0.16$ \\
27 & 03-26 & 3 & 5.2 & 351\,347 & 1.27 & 0.53 & 0.10 & 6.94 & $-0.23$ \\
28 & 03-26 & 4 & 5.0 & 286\,843 & 1.52 & 0.57 & 0.12 & 7.22 & $-0.16$ \\
29 & 03-26 & 4 & 5.0 & 363\,482 & 1.19 & 0.43 & 0.09 & 5.41 & $-0.20$ \\
30 & 03-26 & 4 & 4.1 & 212\,561 & 1.68 & 0.60 & 0.13 & 7.64 & $-0.26$ \\
31 & 03-26 & 4 & 4.2 & 246\,522 & 1.47 & 0.57 & 0.10 & 8.06 & $-0.15$ \\
32 & 04-18 & 1 & 1.3 & 73\,693 & 1.56 & 0.62 & 0.18 & 5.44 & $-0.11$ \\
33 & 04-18 & 1 & 6.5 & 333\,881 & 1.67 & 0.62 & 0.15 & 6.88 & $-0.26$ \\
34 & 04-18 & 1 & 4.9 & 284\,585 & 1.48 & 0.56 & 0.10 & 8.12 & $-0.25$ \\
35 & 04-18 & 1 & 4.2 & 210\,159 & 1.71 & 0.62 & 0.14 & 7.43 & $-0.23$ \\
36 & 04-18 & 2 & 3.4 & 199\,729 & 1.46 & 0.59 & 0.11 & 7.93 & $-0.14$ \\
37 & 04-18 & 2 & 4.2 & 258\,578 & 1.41 & 0.59 & 0.16 & 5.16 & $-0.27$ \\
38 & 04-18 & 2 & 3.0 & 215\,035 & 1.21 & 0.47 & 0.10 & 5.76 & $-0.09$ \\
39 & 04-18 & 2 & 5.2 & 262\,675 & 1.70 & 0.59 & 0.15 & 6.87 & $-0.32$ \\
40 & 04-18 & 3 & 6.0 & 218\,551 & 2.37 & 0.73 & 0.16 & 10.46 & $-0.36$ \\
41 & 04-18 & 3 & 4.8 & 236\,393 & 1.77 & 0.61 & 0.13 & 8.29 & $-0.26$ \\
42 & 04-18 & 3 & 4.8 & 256\,197 & 1.62 & 0.60 & 0.13 & 7.52 & $-0.27$ \\
43 & 04-18 & 3 & 3.7 & 228\,231 & 1.40 & 0.49 & 0.11 & 6.15 & $-0.19$ \\
44 & 04-18 & 4 & 5.3 & 271\,608 & 1.67 & 0.61 & 0.13 & 7.56 & $-0.37$ \\
45 & 04-18 & 4 & 6.8 & 310\,335 & 1.89 & 0.65 & 0.14 & 8.51 & $-0.34$ \\
46 & 04-18 & 4 & 2.0 & 129\,516 & 1.33 & 0.59 & 0.14 & 5.52 & $-0.07$ \\
47 & 04-18 & 4 & 4.1 & 268\,135 & 1.32 & 0.51 & 0.08 & 8.10 & $-0.14$ \\
\end{tabular}
\end{ruledtabular}
\end{table*}

\section{Numerical methods}
\label{sm:num}

Every prediction uses only mean dwells and reward rates measured in fixed-affordance runs
without reward, and it is compared against a separate simulation of the dynamics under
win-stay/lose-sample resampling, in which the Gibbs weight never enters.
Table~\ref{tab:sm-params} lists the parameters.

\subsection{Box}
We integrate Eq.~\eqref{eq:box} on $\D=[-L,L]^2$ with an absorbing wall and reinjection near the
centre. The rotational part of the vector field is integrated \emph{exactly}, by Strang
splitting of rotation against drift and noise, because explicit Euler multiplies the orbit
radius by $\sqrt{1+(a\omega\,\rd t)^2}$ at each step. That spurious outward spiral of
$\sim\!12\%$ per orbit shortens the circular dwell by $\sim\!30\%$ and corrupts $\zeta$, which it
returns as $0.61$ in place of $1$. With exact rotation the circular dwell is $m=9.22$, against
the analytic two-dimensional Bessel first-passage value $L^2/2\sigma^2=8.68$. The exponents
$\zeta=0.964$ (circling) and $0.496$ (drift) come from regressing $\log m$ on $\log\vol$ across
three box sizes. Reward for reaching the wall enters through the rate of wall contact
$\Psi(a)=\E_a\big[(\tau+D_0)^{-1}\big]$, where the offset $D_0$ keeps it bounded on short
excursions, and the equation of state then gives $\Pi\vol/T$ falling from $1.07$ at $\beta=0$ to
$0.73$ at $\beta=8$. For coexistence, the two end modes alone give $\beta^\ast=3.62$ by
Eq.~\eqref{eq:sm-coex}, while summing the tilt $m\,e^{\beta\Psi}$ over all modes places equal
occupation on circling ($a>1/2$) and drift ($a<1/2$) at $3.85$. Simulating the reward rule with
retention $1-e^{-\beta\Psi(a)}$ gives this crossing at $3.89\pm0.04$. If retention instead
depends on the reward $\varrho=(\tau+D_0)^{-1}$ of each excursion, as $1-e^{-\beta\varrho}$,
Eq.~\eqref{eq:sm-mech} predicts $4.05$ and the simulation gives $4.05$.

\subsection{Decision process}
A one-dimensional accumulator $\rd X=\nu\,\rd t+\sigma\,\rd W$ runs between absorbing thresholds
$\pm c$ with reinjection at $0$. Across fifteen thresholds spanning $u=\nu c/\sigma^2=0.10$ to
$15.6$ the simulation reproduces the closed-form mean first-passage time to better than $0.5\%$.
The measured $\zeta$ follows Eq.~\eqref{eq:sm-zetac} to within $0.016$, from $\zeta=2.00$ at the
diffusive end to $1.004$ at the drift end, and the resulting pressure $\Pi c=0.0579$ matches
$T=\sigma^2=0.0576$ to $0.4\%$.

\subsection{Braitenberg vehicle}
A differential-drive vehicle with pose $(x,y,\theta)$ and two light sensors at $\pm\phi$ moves in
a square arena with a light at its centre. With crossing fraction $a$ the wheel speeds are
$v_{L}=v_0+g[(1-a)s_L+a\,s_R]$ and $v_{R}=v_0+g[(1-a)s_R+a\,s_L]$, so that $a<\tfrac12$ steers away
from the light and $a>\tfrac12$ toward it, while the mean speed does not depend on $a$. Noise
enters on the wheels, so translational diffusion is emergent, and the wiring $a$ is redrawn
independently at each wall contact. Over $2.1\times10^5$ rewirings the measured time occupation
matches $m\lambda/Z_0$, with $m$ taken from separate fixed-wiring runs, to a total-variation
distance of $1.1\times10^{-3}$, and the trapped wiring holds $46.3\%$ of the record against
$46.4\%$ predicted from a $10.5$-fold dwell contrast. Fitting $\zeta$ on pairs of arena sizes
across $L=0.5$ to $2.8$ gives $\zeta\approx0.48$ for the ballistic wirings and $\zeta=1.13$ for
the trapped wiring, whose excess over $1$ reflects that the trapped label describes the mode only
asymptotically.

\begin{table}[h]
\caption{Simulation parameters. The box and vehicle are simulated with many parallel realizations
undergoing repeated excursions. The decision process uses the closed-form first-passage law as a
check on the simulated dwells.}
\label{tab:sm-params}
\begin{ruledtabular}
\begin{tabular}{lll}
System & Parameters & Sampling\\
\colrule
Box            & $L=1$, $\omega=6$, $|\mathbf v|=1.6$, $\sigma=0.24$, reward offset $D_0=0.5$, uniform frequency $a\in[0,1]$ & $\sim8\times10^5$ excursions\\
Decision       & thresholds $u=\nu c/\sigma^2\in[0.10,15.6]$, $15$ values & $\ge10^6$ crossings/threshold\\
Vehicle        & sensors at $\pm\phi$, gain $g$, base speed $v_0$, wheel noise $\sigma_w$, $L\in[0.5,2.8]$ & $2.1\times10^5$ rewirings\\
\end{tabular}
\end{ruledtabular}
\end{table}

\end{document}